\documentclass[%
 reprint,
nofootinbib,
 amsmath,amssymb,
 aps,
]{revtex4-1}

\usepackage{graphicx}
\usepackage{dcolumn}
\usepackage{bm}
\usepackage{caption}
\usepackage{subcaption}
\usepackage{float}

\begin{document}

\title{Group Convolutional Neural Networks Improve Quantum State Accuracy}

\author{Christopher Roth and Allan H. MacDonald}
\affiliation{
Physics Department, University of Texas at Austin
}

\begin{abstract}
Neural networks are a promising tool for simulating quantum many body systems. 
Recently, it has been shown that neural network-based models describe
quantum many body systems more accurately when they are constrained to have the 
correct symmetry properties. In this paper, we show how to create maximally expressive 
models for quantum states with specific symmetry properties
by drawing on literature from the machine learning community. 
We implement group equivariant convolutional networks (G-CNN) \cite{cohen2016group}, 
and demonstrate that performance improvements can be achieved
without increasing memory use.   We show that G-CNNs achieve very good accuracy
for Heisenberg quantum spin models in both ordered and spin liquid regimes, and 
improve the ground state accuracy on the triangular lattice over other variational Monte-Carlo methods.
\end{abstract}

\maketitle

\section{Introduction}
Because quantum physics problems with interacting degrees of freedom grow exponentially with 
the number of degrees of freedom, 
accurate numerical solutions are often available only for relatively small system sizes.
Simulations nevertheless play a valuable role in advancing understanding because 
they have the important advantages over experiments on real physical systems 
that the problem being solved is fully characterized and free from unintended disorder. 
Neural network quantum states (NQS) (\cite{carleo2017solving, carrasquilla2020machine,sharir2020deep,hibat2020recurrent,roth2020iterative,nomura2020dirac,liang2021hybrid,astrakhantsev2021broken}) have emerged as a competitive tool for understanding the low temperature properties of quantum many-body 
physics models.  Unlike traditional variational Monte-Carlo (VMC) methods, such as 
Gutzwiller projection,  neural networks have the advantage, and also the disadvantage,
of being free from inductive biases about the structure of the
solution. They compensate for the absence of an informed bias
by using an enormous number of parameters. 
As long as a wide enough model is used, neural networks contain 
arbitrarily accurate solutions \cite{hornik1989multilayer}.  

Although reasonably large neural networks are guaranteed to harbor good solutions in the space of possible parameters, there is no guarantee that these solutions can be found in a reasonable time. 
One way to accelerate training is to constrain the search space by removing non-solutions. This is especially useful on lattice problems which have highly symmetric low-lying energy levels. Recent research \cite{choo2019two,nomura2020dirac,ferrari2019neural,luo2021gauge,luo2020gauge,vieijra2020restricted} has shown that forcing neural networks to have the correct symmetry tremendously improves their ability to model the ground state and low-lying excited 
states of quantum many-body systems.

Other research in the field of NQS has focused on training deeper models, specifically convolutional neural networks (CNN) \cite{broecker2017machine,choo2019two,liang2021hybrid}.
This is motivated by the field of computer vision \cite{he2016deep, szegedy2017inception}, in which deep convolutional neural networks have excelled at pattern recognition. While both images and quantum ground states often have translational symmetry, quantum ground states frequently have additional point group symmetry. In this paper we leverage literature from the machine learning community \cite{cohen2016group} to generalize convolutional NQS to the full wallpaper group of highly symmetric lattices. 

We show that group equivariant convolutional networks (G-CNN) \cite{cohen2016group} provide the most expressive perceptron-like model for wavefunctions with particular symmetry over discrete groups. G-CNNs are constructed using equivariant convolutions, which are the most complex linear operations that preserve
the structure of a discrete symmetry group \cite{cohen2016group}. This approach contrasts with previous research \cite{choo2019two,nomura2020dirac,ferrari2019neural}, which has utilized a symmetry-averaging procedure in which the model is applied to symmetry-transformations of the input and the output is averaged. We show that G-CNNs can be mapped to symmetry-averaged linear 
models by zeroing out some filters between hidden layers. 
When we mask a G-CNN in this manner,
{\it i.e.} when we constrain it to
match previous symmetry-averaging procedures, the performance degrades.

We use a G-CNN with local translational filters to simulate frustrated $J_1-J_2$ Heisenberg models on square and triangular lattices. Both of these Hamiltonians give rise to rich phase diagrams that include magnetically ordered states and spin liquid states that do not order but have non-trivial quantum entanglement. Our model gives accurate results for both triangular and square lattice
phase-diagrams, and compares favorably 
with state-of-the-art variational Monte-Carlo (VMC) results on the triangular lattice.






\section{VMC with Neural Network Quantum States}

We compute ground state energies for the frustrated $J_1-J_2$ Heisenberg model on the square and triangular lattices. The Hamiltonian is given by: 
\begin{eqnarray}
H = J_1 \sum_{i,j \in \langle \rangle} {\bf \sigma}_i \cdot {\bf \sigma}_j + J_2 \sum_{i,j \in
\langle \langle \rangle \rangle} {\bf \sigma}_i \cdot {\bf \sigma}_j,
\end{eqnarray}
where $\langle \rangle$ and $\langle \langle \rangle \rangle$ 
are nearest and next-nearest neighbor links and $J_1,J_2 > 0$. 
We use a neural-network ansatz that associates a complex number $\psi({\bf \sigma})$ with each 
configuration of spins ${\bf \sigma}$. These define the wavefunction:
\begin{eqnarray}
|\psi> = \sum_{\sigma} \psi(\sigma) |\sigma>.
\end{eqnarray}
We optimize our Ansatz using gradient based methods as detailed in Carleo {\it et al.} \cite{carleo2017solving}. 


\section{Symmetric Neural Network Wavefunctions}

Symmetry-averaging neural networks has been shown to improve performance in both machine learning applications \cite{cohen2016group,cohen2019gauge,dieleman2016exploiting,cohen2018spherical,zaheer2017deep} and quantum many-body VMC applications \cite{choo2018symmetries,nomura2020helping,ferrari2019neural}. Below, we first discuss symmetry-averaging from the point of view of equivariance. 
We then show that that G-CNNs provide a richer model class, 
which can be mapped down to symmetry-averaged models by masking filters. 

\subsection{The Principle of Equivariance}

An equivariant function acts on a G-space \cite{cohen2016group,pitts2013nominal}, a set of objects, $S$, that are related by the symmetry transformations of a group, $G$. 
In this paper we consider the set of transformations, generated by lattice
translations and the d4(6) point group for square (triangular) lattices. 
These are the groups of transformations that leave the lattices unchanged. An equivariant function is one that transforms a G-set of poses while preserving the group structure,
\begin{eqnarray}
g'{\cal F}(\{ {\bf f} \}_S) = {\cal F}(g \{ {\bf f} \}_S),
\end{eqnarray}
where $\{ {\bf f} \}_S$ is a set of features over the set $S$. Here $g'$ and $g$ do not be the same operator, but they must be generators of isomorphic groups. For example if $g = T_{x,y}$ is a translation on a periodic 2D square lattice of length L, then $g' = T_{x',y'}$ must also be a translation that is cyclic over L in both dimensions.   

\subsection{Symmetry-Averaged Models are Equivariant Models} \label{sec: sym avg}
Symmetry can be imposed \cite{nomura2020helping} on a model wavefunction $\psi$ 
by a symmetry-averaging procedure that 
applies the model to all symmetry-transformations of the input and 
performs a phase-factor weighted average over outputs:
\begin{eqnarray}  \label{eqn: old sym}
\psi({\bf \sigma}) = \sum_g \chi_{g} {\tilde \psi} (g^{-1} {\bf \sigma}).  
\end{eqnarray}
Here $\chi_{g}$ is the character of symmetry operation $g$, and $g^{-1}$ is the inverse of g, which means that $g g^{-1} = g^{-1} g = I$. This procedure can also be thought of as averaging over the 
output of an equivariant model $\psi_g({\bf \sigma})$ if we define:  
\begin{eqnarray} \label{eqn: sym to equivariant}
\psi_{g}({\bf \sigma})  = {\tilde \psi}(g^{-1} {\bf \sigma}). 
\end{eqnarray}
This definition recasts a single model ${\tilde \psi}$ applied to $N_{G}$ inputs, 
as an equivariant model $\psi_g$ with output dimension $N_{G}$ 
applied to a single input.  Note that
a symmetry transformation on the input to $\psi_g$ yields a symmetry operation on the output: 
\begin{eqnarray}
\psi_g(u^{-1} {\bf \sigma}) = \tilde{\psi}(g^{-1} u^{-1} {\bf \sigma}). = \psi_{u g}({\bf \sigma}).  
\end{eqnarray} 
We see that  is still equivariant, as the set of transformations $\{u\}$ and $\{u^{-1} \}$ form isomorphic groups. Using this new definition, the symmetrization procedure expresses the wavefunction as follows:
\begin{eqnarray}  \label{eqn: new sym}
\psi({\bf \sigma}) = \sum_g \chi_g \psi_g ( {\bf \sigma}).  
\end{eqnarray}
This still produces a $\psi$ that has the desired character:
\begin{eqnarray} 
\begin{aligned} 
\psi(u{\bf \sigma}) = \sum_g \chi_g \psi_g (u{\bf \sigma}) = \sum_g \chi_g \psi_{u^{-1}g} ({\bf \sigma}) \\ = \sum_g \chi_{ug} \psi_g ({\bf \sigma}) = \sum_g \chi_{u} \chi_g \psi_g ({\bf \sigma}) = \chi_{u} \psi({\bf\sigma}). 
\end{aligned}
\end{eqnarray}

From this discussion we see that symmetry-averaging is not 
required to restrict symmetry eigenvalues. Only equivariance is needed. 
In this paper, we replace the symmetry 
averaging procedure by a G-CNN, which consists of stacked equivarant convolutional layers \cite{cohen2016group} interspersed with non-linearities. 
The G-CNN architecture allows us to scale up the number of 
parameters without using more memory, and improves performance on difficult optimization problems. 



\subsection{Generalizing Translational Convolutions to Discrete Groups}

\begin{figure*}
\centering
\begin{subfigure}{.3\textwidth} 
  \centering
  \includegraphics[width=1.\linewidth]{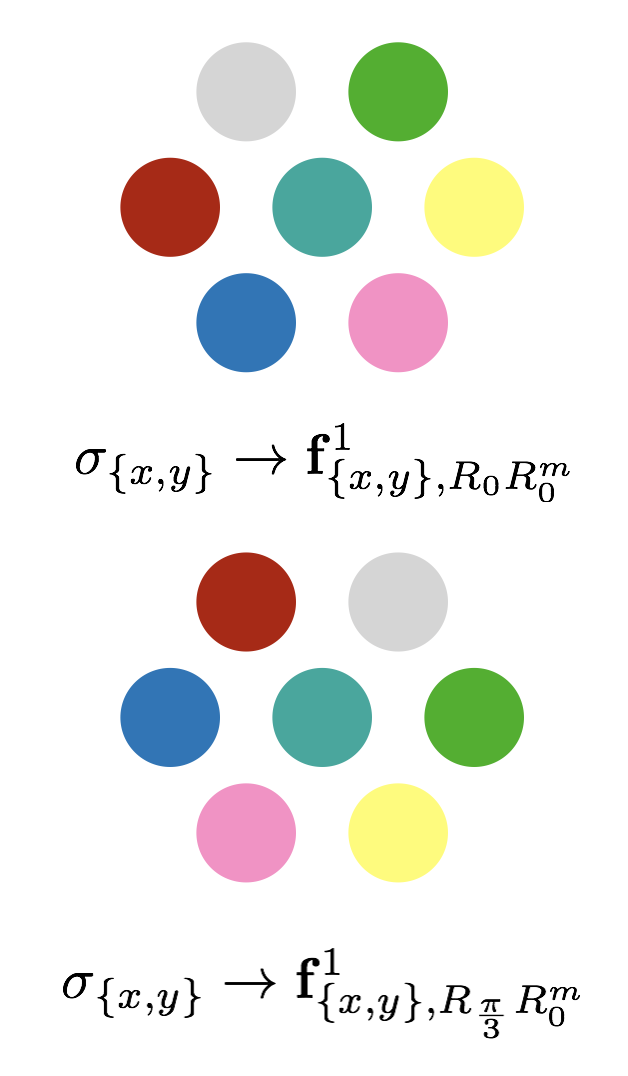}
  \caption{}
    \label{fig: inp to feat}
\end{subfigure}%
\begin{subfigure}{.7\textwidth} 
  \centering
  \includegraphics[width=1.\linewidth]{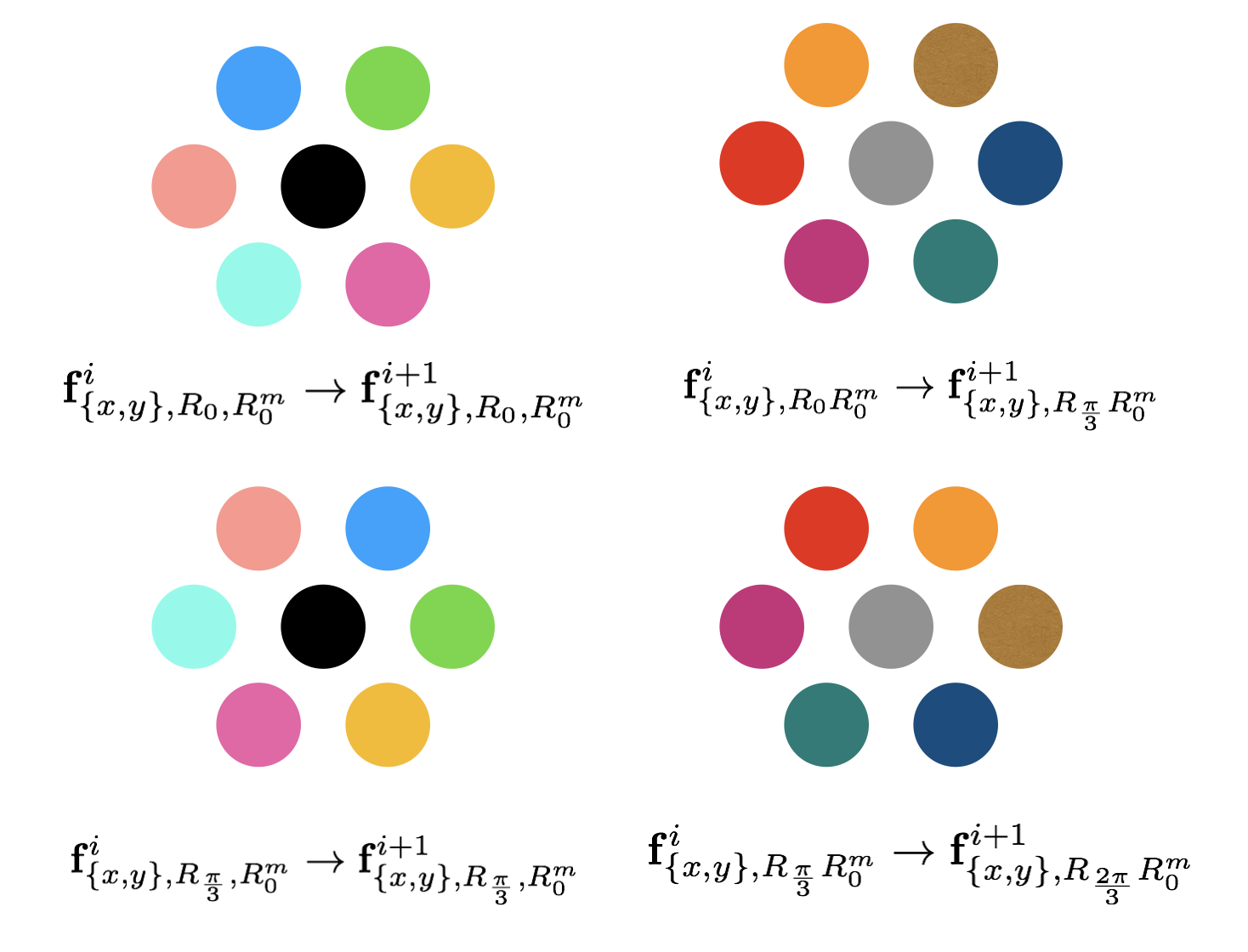}
  \caption{}
    \label{fig: feat to feat}
\end{subfigure}
\caption{Schematic of an equivariant convolutional model on a triangular lattice with p6m symmetry. Nearest neighbor convolutions are drawn for legibility. a) Equivariant convolution connecting the input to the feature maps. Each feature map uses a set of filters with a
specified symmetry transformation over $d_6$. (b) Equivariant convolution connecting feature maps at adjacent layers. 
There are 12 sets of filters that connect poses 
based on their relative orientations.}
\end{figure*}

Equivariance under the translational group is present in 
standard convolutional neural networks (CNNs),
\begin{eqnarray} \label{eqn: trans conv}
C^i_{x,y} = \sum_{x',y' < L} {\bf W}_{x'-x,y'-y}^i \cdot  {\bf f}_{x',y'}, 
\end{eqnarray}
 that satisfy periodic boundary conditions.  The filters ${\bf W}_{x'-x,y'-y}^i$, look at patterns in the feature map ${\bf f}_{x',y'}$ separated from ${x,y}$ by 
$\{x'-x,y'-y\}$. CNNs work by alternating convolution operations with non-linearities to develop increasingly abstract representations of the input. 

The G-CNN generalizes the CNN to act over a discrete group $G$, which may contain non-commuting operations. For our implementation, we consider the full wallpaper group, which adds rotation and mirror symmetry on top of translation. The group convolution operation is written as follows: 
\begin{eqnarray} \label{eqn: gen conv}
C^i_g= \sum_{h \in G} {\bf W}_{g^{-1} h}^i \cdot  {\bf f}_h. 
\end{eqnarray}
For the G-CNN, the feature map is defined over the G-space generated by the wallpaper group, instead of just the translation group. We see that the group convolution is equivariant,
since a group operation on input yields the same group operation on the output:
\begin{eqnarray}
\sum_{h \in G} {\bf W}^i_{g^{-1} h} \cdot  {\bf f}_{uh} =  \sum_{h \in G} {\bf W}^i_{g^{-1} u^{-1} h} \cdot  {\bf f}_{h} = C^i_{ug}.
\end{eqnarray} 
This linear operation is the building block of our model.  


\subsection{Our Model}
\begin{figure*} 
\centering
\begin{subfigure}{.4\textwidth}
  \centering
  \includegraphics[width=1.\linewidth]{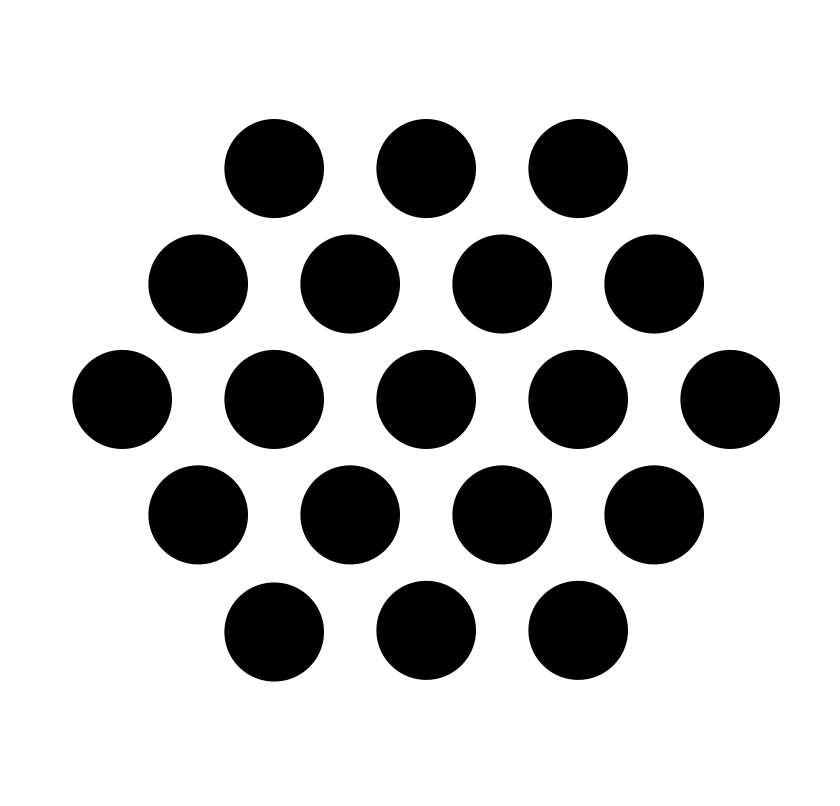}
  \caption{}
\end{subfigure}%
\begin{subfigure}{.4\textwidth}
  \centering
  \includegraphics[width=1.\linewidth]{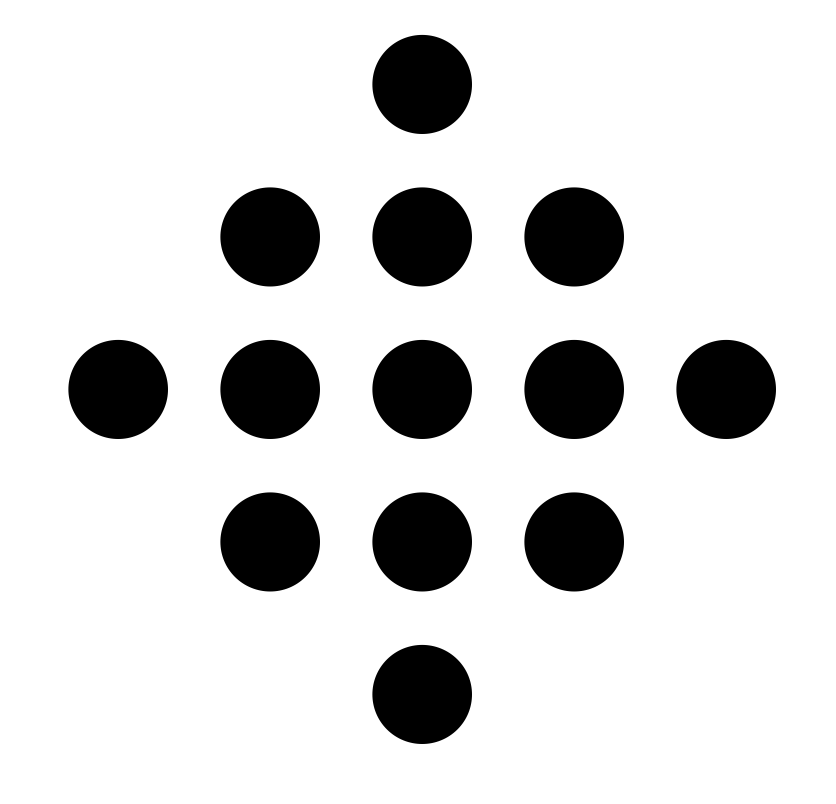}
  \caption{}
\end{subfigure}
\caption{Filter shapes for our restricted equivariant convolution operation on the a) triangular lattice  (b) square lattice. For both lattices, interactions are 
restricted to third nearest neighbors.}
\label{fig: filter shapes}
\end{figure*}

We use a model that is similar to a G-CNN \cite{cohen2016group}, and
specified by a stack of equivariant convolutional layers interspersed with pointwise non-linearities. The first convolution takes the z-components of the spins, $\sigma_{\bf x}$, where ${\bf x}$ are the lattice site labels, 
and outputs a feature map over the full wallpaper group: 
\begin{eqnarray} \label{eqn: input layer}
{\bf f}^{1}_g  = \Gamma \Big( \sum_{{\bf x}} W_{g^{-1} {\bf x}}^0  \sigma_{\bf x} \Big),
\end{eqnarray}
where $\Gamma$ is a point-wise non-linearity. 
This convolution is diagrammed in Fig. \ref{fig: inp to feat}. The feature-to-feature convolutions are performed as in equation \ref{eqn: gen conv}:
\begin{eqnarray} \label{eqn: layer}
{\bf f}^{i+1}_g  = \Gamma \Big( \sum_{h \in G} W_{g^{-1} h}^i  {\bf f}^i_h \Big).
\end{eqnarray}
The structure of these convolutions is seen in Fig. \ref{fig: feat to feat}.
We compute our wavefunction with character $\{\chi_g \}$ by phase-weighting over the exponential our final embedding, $f_g^N$, which has length one:
\begin{eqnarray}   
\psi({\bf \sigma}) = \sum_g \chi_{g^{-1}} \textrm{exp}(f_g^N).
\end{eqnarray}
We choose $\Gamma$ to be the SELU nonlinearity \cite{klambauer2017self} applied separately to the real and imaginary parts, 
\begin{eqnarray} 
\Gamma(x) = \textrm{SELU}(\textrm{Re}(x)) + i \textrm{SELU}(\textrm{Im}(x)).
\end{eqnarray}
The SELU nonlineariity moves the distribution of activations in the direction of zero mean and one variance, which enables stable training of deep networks. We note that for all models studied in this paper, the ground state is fully symmetric, i.e. $\chi_g = 1$ for all g.

While this model may appear complicated, we believe it is actually quite simple. It is just a multi-layer perceptron, with alternating linear and pointwise non-linear transformations, in which weights are tied together to force particular symmetry properties. Training a simple model, with a constant feature dimension and with neither bells nor whistles, will demonstrate the power of this approach. 



\section{Results}

We begin by showing results from our best performing model, which uses a G-CNN architecture with local filters. This model achieves state-of-the-art VMC ground state energies on the triangular lattice,
and competitive energies on the square lattice. 
We then examine how the perfomance of the G-CNN is affected by masking off-diagonal filters. As detailed in section \ref{sec: mapping to others}, this 
simplification maps G-CNNs to models that are forced to be equivariant by symmetry-averaging. The optimization details for both of these experiments are outlined in section \ref{sec: opt details}.   

\subsection{Results from our Best Performing Model} 

In Table \ref{table: Triangular} we summarize a variety of 
ground state energy estimates obtained using different methods for 6$\times$6 finite-size $J_1,J_2$ 
triangular lattice Heisenberg models.  For this lattice size, the 
DMRG\cite{iqbal2016spin} and and ED\cite{bernu1994exact} estimates
agree to five figures and are essentially exact.  For the $J_2=0$ case
our model achieves 
$\sim 0.2\%$ accuracy for the energy, compared to the $\sim 1\%$ accuracy 
achieved by other variational methods in the literature.
For $J_2=J_1/8$, in the spin-liquid regime, the ground state 
energy error is reduced by a factor of $\sim 3$ compared to literature VMC estimates.
VMC calculations often add Lanczos steps \cite{sorella2001generalized} to
improve variational energy accuracies;  
our ground state energies surpasses all other VMC methods in accuracy
even without this elaboration.

\begin{figure*}
\centering
  \includegraphics[width=.9\linewidth]{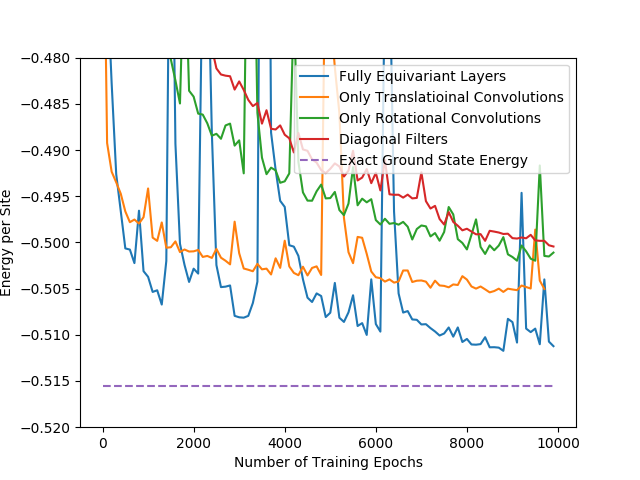}
  \caption{Performance of the G-CNN on the triangular Heisenberg model ($J_2 = \frac{1}{8}$) as various off-diagonal filters are set to zero. As described in section \ref{sec: mapping to others} this is equivalent to replacing group convolution with symmetry averaging. All models are trained using the Adam optimizer with a learning rate of $3 \times 10^{-3}$ }
    \label{fig: compare}
\end{figure*}

\begin{table}[h] 
\begin{center} 
\begin{tabular}{ |c|c|c| } 
 \hline
 & $J_2 = 0$ & $J_2 = J_1/8$ \\ \hline
ED \cite{bernu1994exact} & -0.5603734 & -0.515564 \\ \hline
DMRG \cite{iqbal2016spin} & -0.560375 &  -0.51557 \\ \hline
G-CNN & -0.55922 & -0.51365  \\ \hline
NN + Gutzwiller \cite{ferrari2019neural} & -0.553 & N/A \\ \hline
VMC + 2LS \cite{iqbal2016spin} & N/A & -0.512503 \\ \hline
VMC \cite{kaneko2014gapless} & -0.55519 & -0.5089 \\ \hline
VMC \cite{mezzacapo2010ground} & -0.55420 & N/A \\ \hline
VMC \cite{iqbal2016spin} & -0.548025 & -0.501788 \\ \hline
\end{tabular}
\end{center}
 \caption{Comparison between finite size $6 \times 6$ triangular lattice 
 ground state energies estimates}
 \label{table: Triangular}
\end{table}

We studied the dependence of performance, measured by the accuracy of 
ground state energy estimates, on the range of the filter for $J_2=J_1/8$. 
Restricting the connectivity reduces computational overhead, and biases the model towards learning short range interactions.  Changing the filter shape changes the way in which the model conveys information about the connectivity of the lattice.  We tried a few different symmetric filter shapes and settled on a version that convolves over third-nearest-neighbors,
as shown in Fig.~\ref{fig: filter shapes}.

In order to demonstrate the robustness of our model, we applied the same 
architecture to the ordered and spin liquid regimes of the square lattice $J_1-J_2$ Heisenberg model. Remarkably, across all four domains, we only had to adjust a single learning rate. The results on the square lattice are summarized in Table \ref{table: Square}


\begin{table}[h]
\begin{center} 
\begin{tabular}{ |c|c|c|c| } 
 \hline
 & G-CNN & CNN + $c_4$ & RBM + PP \\ \hline
$J_2 = J_1/2$ & $0.17\%$ & $0.39 \%$ & $0.022\%$ \\ \hline
$J_2 = 0$ & $0.016 \% $ & $0.007\%$ &  N/A \\ \hline
Max Memory & $16 \times N_g$ & $10 \times N_g$ & $576 \times N_g$ \\ \hline
$N_{params}$ & 113360 & 3838 & 22032 \\ \hline
\end{tabular}
\end{center}
 \caption{Errors relative to exact diagnalization for $6 \times 6$ square-lattice finite-size 
 ground state states. The number of variational parameters and the memory usage are compared for 
 different approaches.}
 \label{table: Square}
\end{table}

We make a direct comparison to the square lattice CNN model described in Choo et al. \cite{choo2019two} who use a translationally equivariant CNN symmetry-averaged over $c_4$, while we use a model that is equivariant over the full space group $p4m$. 
As explained in section \ref{sec: mapping to others} of the appendix, their model can be mapped to a G-CNN over $p4$ with masked off-diagonal rotational filters. Our model uses slightly more memory, while their model has more layers and 
uses an improved optimizer. We see that the G-CNN model 
nevertheless has better ground state energy accuracy on the difficult-to-simulate spin-liquid state. 
Neither CNN architecture is competetive with the architecture detailed by Nomura et al. \cite{nomura2020dirac}, who combine a restricted Boltzmann machine with a pair product state (RBM+PP). Theirs is a shallower model that uses far more memory. 
For the ordered state, both CNN based methods are very accurate, and the discrepancy may be due to convergence issues \cite{reddi2019convergence} with the adaptive moment (Adam) 
parameter optimizer \cite{kingma2014adam}. 


We note that the equivariant model has by far the best parameter-number to
memory ratio, as it adds connectivity along the entire $d_4$ symmetry group. 
Maximizing the number of parameters in a memory profile \cite{kitaev2020reformer,fedus2021switch} is important  for performance improvement on 
massively parallel modern hardware. 
Even though our model has more parameters than the RBM + PP model, it is likely much faster to train on a GPU. 



\subsection{Effect of Off-Diagonal Filters}
We document the importance of full connectivity using the $J_1-J_2$ Heisenberg model on a triangular lattice with $J_2 = J_1/8$. 
This is a difficult system to simulate since it gives rise to a spin liquid state with no magnetic order and non-trivial entanglement. We train four models with different feature-to-feature connectivities 
to test the effect of replacing layer-wise equivariance with symmetry averaging. We do this by applying a masking procedure, where we set off-diagonal filters to zero. Here, the term off-diagonal distinguishes between filters that connect poses of the same orientation in a particular symmetry group and filters that connect different orientations. As an example, filters in equation \ref{eqn: trans conv} with $\{x',y'\} = \{x, y\}$ are translationally diagonal whereas all others are translationally off-diagonal.  

First we train a model with diagonal feature-to-feature interactions along all groups, corresponding to a fully connected linear model symmetrized over $p6m$. Second we train models with off-diagonal features in translation and rotation set to zero. These correspond to models that convolve over some degrees of freedom, and symmetry average over others.   
Finally, we compare all of these with a G-CNN  with full connectivity. While all four of these models use the same amount of memory, they have vastly different numbers of parameters.
\begin{table}[h]
\begin{center} 
\begin{tabular}{ |c|c|c|c|c| } 
 \hline
 Connectivity & Full & Translational & $d_6$ & Diagonal  \\ \hline
$N_{params}$ & 449856 & 38016 & 13056 & 1616 \\ \hline
\end{tabular}
\end{center}
 \caption{Comparison between the number of parameters of different equivariant architectures}
 \label{table:2dheisenberg}
\end{table}

We see in Fig. \ref{fig: compare} that adding a richer layer structure systematically improves the performance. The model with fully-connected layers markedly outperforms the others, surpassing their performance in $1/10$ of the training time.  We also see in Fig. \ref{fig: compare} that the energy estimates are not monotonic functions of 
training time.  This behavior is related to our use of high learning rates that enable escapes from 
local minima. Although higher learning rates worsen the performance temporarily, we find that the final variational energies are ultimately superior.   

We see that translational filters have a bigger effect on performance than rotational filters. This could simply be due to the number of parameters in the models. On a $6 \times 6$ lattice, adding translational interactions increases
the number of parameters by a factor of $36$, while adding $d_6$ interactions only increases
the number of parameters by a factor of $12$.

\section{Discussion and Conclusion}

In this paper we use a G-CNN to find the ground wavefunction of the frustrated Heisenberg model on the square and triangular lattices. We find that the G-CNN architecture yields accurate ground state energies for both magnetically ordered states and spin liquids, including the most accurate VMC energies on the triangular lattice to date. 

Among perceptron-like models, G-CNN models are more expressive
than symmetry-averaged models. We explicitly 
demonstrate how performance degrades as the model class is constrained. 
For the wallpaper group $p6m$, we see that it is important to convolve over 
both translation operations and the point group. 


For our experiments we used the simplest kind of neural network model 
that satisfies the symmetry constraints, a multi-layer perceptron with tied weights. 
We note that other advances in machine learning, such as depthwise convolutions \cite{chollet2017xception} and attention \cite{vaswani2017attention}, can also be implemented with group equivariance \cite{romero2020attentive}. 
Implementing these models could certainly further improve the performance of the quantum spin model ground state 
calculations we have discussed, and other quantum ground state calclations.
Although we only studied relatively small systems, these calculations could be scaled to larger systems
by combining sparsification, transfer learning,
and efficient sampling \cite{yang2020scalable}. 

Finally, we note that several other papers have studied the importance of using equivariant models applied to different symmetry groups. Luo {\it et al.} \cite{luo2020gauge} show that gauge equivarant neural networks improve performance on gauge invariant Hamiltonians \cite{luo2021gauge}, and Pfau {\it et al} \cite{pfau2020ab} use  a permutation 
equivariant model to account for identical electrons.  Since many important physics models have substantial symmetry,
applications of equivariant models are likely to be important for machine-learning methods in computational physics.  


\section{Code}
Code for implementing G-CNNs on NQS will be made publically available through NetKet \cite{carleo2019netket} in the near future. Pytorch code for these experiments will be uploaded at https://github.com/chrisrothUT/

\section{Acknowledgements} Chris Roth acknowledges helpful interactions with Mohamed Hibat-Allah, Juan Carrasquilla and Giuseppe Carleo.  This work was supported in part by 
the Army Research Office (ARO) Grant $\#$ W911NF-17-1-0312 (MURI).

\section{Appendix} 
\subsection{Implementation of the Equivariant Convolution}
We implement our equivariant convolution exactly as in Cohen {\it et al.} \cite{cohen2016group}. We learn an index mapping that relates our filters of shape $[d_{in},d_{out},N_g]$ to a group of 2D convolutions with rotated filters as depicted in \ref{fig: feat to feat}. Then we concatenate the rotation/reflection with the channel dimension and apply a 2D convolution. For details see the aforementioned paper.

\subsection{Equivariant Models can be Mapped to other Symmetrized models by Constraining Filters} \label{sec: mapping to others}

Recently several groups \cite{nomura2020dirac,choo2019two,ferrari2019neural,szabo2020neural,liang2021hybrid} have accurately simulated the frustrated $J_1-J_2$ Heisenberg model on a square lattice by constraining the symmetry eigenvalues. While some of these models used translational convolutions, none of them convolved over the entire wallpaper group $p4m$ as done by the G-CNN. 


We demonstrate how the G-CNN can be mapped down to other multi-layer linear models by masking filters. For the sake a brevity, we show the mapping between the p4m G-CNN and a) a linear model symmetry-averaged over p4m b) a translationally equivariant model symmetry averaged over $d_4$.    


We construct a symmetry operation in $p4m$ by first applying a reflection, followed by a rotation and a translation, 
\begin{eqnarray}
S_{x,y,\alpha,\beta} = T_{xy} R_\alpha R^m_\beta.
\end{eqnarray}
Using equation \ref{eqn: gen conv}, we can write the filter that connects two G-set elements in neighboring layers. 
\begin{eqnarray}
S^{-1}_{x,y,\alpha,\beta} S_{x',y',\alpha',\beta'} = R^m_{-\beta} R_{-\alpha} T_{x'-x,y'-y} R_{\alpha'} R^m_{\beta'}.
\end{eqnarray}
This is a convolution operation with filters transformed by $p4m$ as shown in Fig. \ref{fig: feat to feat}.

To map to a multi-layer linear model symmetry-averaged over $p4m$, the input-to-feature filters are kept the same and the feature-to-feature filters are restricted to be diagonal, $\alpha = \alpha'$, $\beta = \beta'$, $x = x'$ and $y = y'$. 

The new feature-to-feature convolution becomes
\begin{eqnarray} \label{eqn: layer}
{\bf f}^{i+1}_{x,y,\alpha,\beta} = \sigma\Big(W^i {\bf f}^i_{x,y,\alpha,\beta} \Big).
\end{eqnarray}
This is equivalent to applying the same linear model to each pose. The different poses are generated by the symmetry transformed filters shown in Fig. \ref{fig: inp to feat}. Equivalently, we could generate the poses by symmetry transforming the input and keeping the filters the same. As a result, we see that the  G-CNN with diagonal feature-to-feature filters is equivalent to a linear model symmetrized over $p4m$.  


We can also map the G-CNN to a translationally equivariant model (CNN) symmetrized over d$_4$ by constraining $\alpha = \alpha'$  and $\beta = \beta'$. With this constraint the feature-to-feature convolution is 
\begin{eqnarray} 
{\bf f}^{i+1}_{x,y,\alpha,\beta} = \sum_{x', y'} \sigma \Big(W^i_{x'-x,y'-y,\alpha,\beta} {\bf f}^i_{x',y',\alpha,\beta} \Big).
\end{eqnarray}
The feature-to-feature convolution now consists of CNNs applied to each pose in $d_4$ labeled by $\alpha, \beta$ where the filters of each CNN are transformed according to the group element
\begin{eqnarray}
W^i_{x'-x,y'-y,\alpha,\beta} = R^m_{-\beta} R_{-\alpha} W^i_{x'-x,y'-y} R_{\alpha} R^m_{\beta}
\end{eqnarray}
Again, if we change basis and transform the input instead of the filters, this is a CNN symmetrized over $d_4$. Both of these examples show that symmetry-averaging creates a more restrictive model class than the G-CNN for multi-layer networks.

\subsection{Triangular Heisenberg on a Rectangular Grid}
We implement the triangular Heisenberg model on a rectangular graph by noting that a triangular graph is isometric to a rectangular graph with coupling along one of the diagonals. We implement our models on $6 \times 6$ tori with periodic boundary conditions (PBCs) and coupling along one of the diagonals. This is equivalent to a hexagonal shaped lattice with $36$ sites.

\subsection{Coordinate Mappings}
On the square lattice, a $\frac{\pi}{2}$ rotation is defined by the coordinate mapping $(i,j )\rightarrow (-j, i)$ where the coordinates are defined modulo L. On the triangular lattice a $\frac{\pi}{3}$ mapping is accomplished by a coordinate mapping $(i,j )\rightarrow (i-j, i)$. We can arbitrarily choose a reflection axis, and for simplicity we do the coordinate transform $(i,j )\rightarrow (j, i)$ for both geometries.

\subsection{Optimization Details} \label{sec: opt details}
Our best performing model had $4$ layers and $16$ hidden nodes. We used the Adam optimizer \cite{kingma2014adam} with a learning rate of $3 \times 10^{-3}$ and the default parameters. We optimized our learning rate and network architecture on the triangular lattice with $J_2 = 0.125$ and used the same hyperparameters for the other systems, except for the square lattice with $J_2$ = 0 where we found that the learning rate was too high, and decreased it by a factor of 3. 

Our optimization procedure was as follows. First, we optimized the phase structure for $500$ steps by setting $Re(log(\psi)) = 0$ before outputting $\psi$. We found that pre-optimizing the phases \cite{szabo2020neural}, which helps avoid local-minima, was crucial to successfully finding the ground state. We then trained both amplitudes and phases for $10^{4}$ steps at batch size $100$ before increasing the batch size to $1000$ and training for an additional $2 \times 10^3$ steps. 

We tuned the learning rate (on the triangular lattice at $J_2 = \frac{1}{8})$ by altering it by factors of 3 and modified N$_{hidden}$ and N$_{layers}$ under the constraint $N_{hidden} \times N_{layers} = 64$ We found that a ratio of $\frac{N_{hidden}}{N_{layers}} = 4$ worked best.






\bibliographystyle{unsrt}
\bibliography{equivariant.bib}

\end{document}